\documentclass[fleqn,usenatbib,onecolumn]{mnras}
\def\paren#1{\left( #1 \right)}

\usepackage[T1]{fontenc}
\usepackage{ae,aecompl}
\usepackage{graphicx}
\usepackage{epstopdf}
\usepackage{amsmath}
\usepackage{amssymb}
\usepackage{enumerate}
\usepackage{color}

\title[Structured Jets as EM Counterparts]{Electromagnetic Counterparts to Structured Jets from Gravitational Wave Detected Mergers}
\author[G.P. Lamb and S. Kobayashi]{Gavin P Lamb$^{1}$ and Shiho Kobayashi$^{1}$
\\
$^{1}$Astrophysics Research Institute, LJMU, IC2, Liverpool Science Park, 146 Brownlow Hill, Liverpool L3 5RF, 
UK}
\date{Accepted XXX. Received YYY; in original form ZZZ}
\pubyear{2017}
\begin{document}
\label{firstpage}
\pagerange{\pageref{firstpage}--\pageref{lastpage}}
\maketitle

\begin{abstract}
We show the peak magnitude for orphan afterglows from the jets of gravitational wave (GW) detected black-hole/neutron star - neutron star (BH/NS-NS) mergers highly depends on the jet half-opening angle $\theta_j$.
Short $\gamma$-ray bursts (GRB) with a homogeneous jet structure and $\theta_j>10^\circ$, the orphan afterglow viewed at the typical inclination for a GW detected event, 38$^\circ$, is brighter at optical frequencies than the comparable macronova emission.
Structured jets, where the energetics and Lorentz factor $\Gamma$ vary with angle from the central axis, may have low-$\Gamma$ components where the prompt emission is suppressed;
GW electromagnetic (EM) counterparts may reveal a population of failed-GRB orphan afterglows.
Using a Monte Carlo method assuming a NS-NS detection limit we show the fraction of GW-EM counterparts from homogeneous, two-component, power-law structured, and Gaussian jets where the variable structure models include a wide low energy and $\Gamma$ component:
for homogeneous jets, with a {$\theta_j=6^\circ$ and typical short GRB parameters, we find {\it r}-band magnitude $m_r\leq21$ counterparts for $\sim 13.6\%$ of GW detected mergers;
where jet structure extends to a half-opening angle of $25^\circ$, two-component jets produce $m_r\leq21$ counterparts in $\sim30\%$ of GW detected mergers;
power-law structured jets result in $\sim37\%$;
and Gaussian jets with our parameters $\sim13\%$.}
We show the features in the lightcurves from orphan afterglows can be used to indicate the presence of extended structure.
\end{abstract}
\begin{keywords}
gamma-ray bursts: general - gravitational waves
\end{keywords}

\section{Introduction}
The merger of binary neutron star (NS) systems or black-hole (BH) neutron star systems are thought to be the progenitors of short gamma-ray bursts (GRB) \citep{1992ApJ...395L..83N,1993Natur.361..236M,2007ARep...51..308B,2007PhR...442..166N,2014ARA&A..52...43B}.
The rapid accretion of a merger debris disc onto a compact object can power relativistic bi-polar jets.
Jet energy is initially dissipated internally producing the prompt $\gamma$-rays of a GRB.
The jet interacts with the ambient medium at later times and develops an external shock which expands and produces a broadband afterglow \citep[e.g.][]{2004RvMP...76.1143P,2004IJMPA..19.2385Z}.

The inspiral and merger of a NS-NS or BH-NS system is caused by the emission of gravitational waves (GW).
Such GWs are a target for ground-based GW detectors such as advanced LIGO, Virgo, and KAGRA \citep{2016LRR....19....1A,2013PhRvD..88d3007A}.
The merger of binary BH systems produced the advanced LIGO detections GW150914, GW151226, GW170104, and the 87\% confidence LVT151012 \citep{2016PhRvX...6d1015A,2017PhRvX...221101}.
BH-BH mergers are not expected to produce an EM counterpart, however see \cite{2016ApJ...826L...6C,2016ApJ...823L...2A,2016ApJ...820L..36S,2017arXiv170600029V};
and various scenarios have been suggested \cite[e.g.][]{2016ApJ...819L..21L,2016ApJ...821L..18P,2016ApJ...827L..31Z,2016PTEP.2016e1E01Y}.
To maximize the science returns from GW astronomy the detection of an EM counterpart is essential; 
GW from NS-NS and BH-NS mergers should be detected within the next few years.
GW detections of BH/NS-NS mergers will trigger a broad-band search for electromagnetic (EM) counterparts.
However, short GRBs rarely occur within the range of GW detectors, $\sim 300$ Mpc for face-on NS-NS mergers \citep{2010CQGra..27q3001A};
this is possibly due to the high collimation of the prompt $\gamma$-ray emission, where $\sim 0.5\%$ of jets with a half-opening angle $\theta_j\sim6^\circ$ would be inclined towards an observer, or a mis-match between short GRB peak energies and the {\it Swift} detection band makes detection more difficult.
However, the afterglows from the merger jets may be observable as `off-axis' orphans.
Alternatively a large fraction of the jets from such mergers may have no bright prompt emission due to a low bulk Lorentz factor \citep{2016ApJ...829..112L}.
More isotropic EM counterparts are often discussed to localize a large sample of GW events \citep[e.g.][]{2011Natur.478...82N,2012ApJ...746...48M,2013ApJ...767..124N,2013ApJ...771...86G,2015MNRAS.446.1115M,2015ApJ...802..119K,2016ApJ...831..190H}.

Other than the bi-polar jets, numerical simulations of NS-NS and BH-NS mergers show sub and mildly relativistic ejecta \citep[e.g.][]{2000A&A...360..171R,2001A&A...380..544R, 2008PhRvD..78f4054Y,2010PhRvL.104n1101K,2012PhRvD..85d4015F,2013ApJ...776...47D, 2013PhRvD..87b4001H,2015PhRvD..91l4041D,2016ApJ...825...52K,2017CQGra..34j5014D, 2017PhRvD..95f3016C}.
Such ejecta is {more} isotropic in the case of a NS-NS merger and {highly anisotropic} for BH-NS mergers \citep{2015PhRvD..92d4028K}.
This merger ejecta can produce macronovae (also called kilonovae) from the decay of r-process nucleosynthesis products \citep[e.g.][]{1998ApJ...507L..59L,2013Natur.500..547T,2013ApJ...774L..23B}.
Macronovae typically peak at red wavelengths with $\ga22$ magnitude for a source at 200 Mpc \citep{2016AdAst2016E...8T}.
Radio flares are expected at much later times; 
1-4 years and $\sim 1$ mJy \citep{2011Natur.478...82N,2016ApJ...831..190H}.
Aditionally, the jet must propagate through the merger ejecta, forming a cocoon that can collimate the outflow \citep{2011ApJ...740..100B,2014ApJ...784L..28N}.
A resultant cocoon-ejecta shock may give rise to X-ray or UV/optical emission \citep{2017ApJ...834...28N,2017MNRAS.471.1652L}.
The jet will break out of the merger ejecta and continue to propagate into the ambient medium where the collimating pressure from the cocoon is lost.
This transition can result in the jet becoming structured i.e. the energy $\epsilon$ and bulk Lorentz factor $\Gamma$ vary across the jet cross-section \citep[e.g.][]{2001ARep...45..236L,2002ApJ...571..876Z,2002MNRAS.332..945R}.
Low-$\Gamma$ components of a structured jet will give rise to EM counterparts to a NS-NS or BH-NS merger without the bright prompt $\gamma$-ray emission.
Given a GW detection from a NS-NS or BH-NS merger, jet external shock EM counterparts will be able to reveal the jet structure.

In \S \ref{structure} we describe the jet structures considered in this paper; 
in \S \ref{method} we give details of the model used to estimate the observable emission at any inclination and show the results of our Monte Carlo;
in \S \ref{disc} we discuss the various afterglow peak flux and peak time distributions;
and in \S \ref{conc} we give concluding remarks and comment on the results implications to EM counterpart searches for GW detected compact stellar mergers.

\section{Jet Structure}
\label{structure}
Jet structure refers to the opening angle and energy distribution within a relativistic jet;
the jets in GRBs are usually assumed to have a simple `top-hat' or homogeneous jet structure where the energy per unit solid angle $\epsilon$ and the bulk Lorentz factor $\Gamma$ are uniform until a sharp edge at the jet opening angle.
Structured jets, where the energy distribution varies with angle from the centre, have been discussed in relation to long GRBs;
The structure is a result of the jet breaking out from the stellar envelope \citep[e.g.][]{2002bjgr.conf..146L, 2003ApJ...594L..19L, 2003ApJ...586..356Z, 2004ApJ...608..365Z, 2005ApJ...629..903L, 2010ApJ...723..267M, 2015MNRAS.447.1911P}.
Alternatively, the structure can be a result of the jet formation mechanism \citep[e.g.][]{2003ApJ...594L..23V, 2003ApJ...584..937V}, an accretion disc forms that can launch a relativistic jet, either by the Blandford-Znajek (BZ) mechanism \citep{1977MNRAS.179..433B} or neutrino annihilation \citep[e.g.][]{1999ApJ...518..356P}.
If the jet from a NS-NS or BH-NS merger propagates through an outflow at early times, then upon break-out some structure can be expected; 
similarly, if the jet is formed and accelerated by either BZ or neutrino annihilation, or a combination of both, then the structure can arise from the various components i.e. spine and sheath.
{Such jet structure could enhance the GW-GRB association probability \citep[e.g.][]{2017arXiv170807008J,2017arXiv170807488K}.}

Other than homogeneous jets, there are three alternative jet structures that are commonly discussed \citep[e.g.][]{2002ApJ...570L..61G,2003A&A...400..415W,2005MNRAS.363.1409P}:
\begin{enumerate}[(i)]

\item A two-component or spine and sheath jet; 
a fast, narrow core and a slower, wider sheath \citep[e.g.][]{2003ApJ...594L..23V, 2005ApJ...626..966P,2007ApJ...656L..57J}. 
Also see \cite{2011MNRAS.417.2161B} where the wider component is faster.
Alternatively, baryon loading of the jet edges where a structured magnetic field prevents charged baryon drift into the jet core, will create a jet with uniform energy but a wider low-$\Gamma$ component \citep{2013ApJ...765..125L}.
The general two-component jet $\epsilon$ and $\Gamma$ follow
\begin{equation}
\epsilon(\theta) = \begin{cases} \epsilon_c \quad &\theta<\theta_c, \\
\epsilon_s \quad &\theta>\theta_c, \end{cases} \qquad \Gamma(\theta) = \begin{cases} \Gamma_c \quad &\theta<\theta_c, \\ \Gamma_s \quad &\theta>\theta_c, \end{cases}
\label{eq_4}
\end{equation}
where the subscript $c$ indicates the jet core parameter, and the subscript $s$ indicates the uniform sheath parameter.

\item A structured jet where the energy and Lorentz factor are a function of the jet angle outside a uniform core \citep[e.g.][]{2002ApJ...571..876Z,2002MNRAS.332..945R,2004MNRAS.354...86R,2003ApJ...591.1075K}.
The jet $\epsilon$ and $\Gamma$ follow
\begin{equation}
\epsilon(\theta) = \begin{cases} \epsilon_c \quad &\theta<\theta_c, \\
\epsilon_c \paren{\frac{\theta}{\theta_c}}^{-k_e} \quad &\theta>\theta_c, \end{cases} \qquad \Gamma(\theta) = \begin{cases} \Gamma_c \quad &\theta<\theta_c, \\ \Gamma_c \paren{\frac{\theta}{\theta_c}}^{-k_\Gamma} \quad &\theta>\theta_c, \end{cases}
\label{eq_2}
\end{equation}
where $\theta$ is the angle from the jet axis, and we assume uniform baryon loading where $k_e=k_\Gamma=k\geq 0$.

\item A Gaussian jet \citep[e.g.][]{ 2003ApJ...591.1075K,2004IJMPA..19.2385Z,2004MNRAS.354...86R}.
The jet $\epsilon$ and $\Gamma$ follow
\begin{equation}
\epsilon(\theta) = \epsilon_c \rm{e}^{-\paren{{\theta^2}/{2\theta_c^2}}}, \qquad \Gamma(\theta) = \Gamma_c \rm{e}^{-\paren{{\theta^2}/{2\theta_c^2}}}. \label{eq_6}
\end{equation}

\end{enumerate}
In all cases $\theta < \theta_j$, where $\theta_j$ is the maximum jet half opening angle.
The existence of a jet edge is motivated by numerical simulations of compact stellar mergers \citep[e.g.][]{2011ApJ...732L...6R} where resistive-magnetohydrodynamics simulations result in a jet-like magnetic structure with a half-opening angle of $\sim25^\circ$ \citep{2015PhRvD..92h4064D}.
The jets are assumed to be symmetric about the central axis.
Observed emission from the various components of a jet depends on the viewing angle $\theta_{\rm obs}$, measured from the jet-axis.

\section{Method and Results}
\label{method}

The jet energy dissipated by internal processes \citep[e.g.][]{1994ApJ...430L..93R,2011ApJ...726...90Z} is radiated as $\gamma$-rays via the synchrotron process.
The radius of this internal dissipation from the central engine can be estimated using the minimum variability of the prompt emission, typically $\delta t \sim 0.1$ s \citep{2002MNRAS.330..920N},
\begin{equation}
R_\gamma \simeq \Gamma^2 c \delta t \simeq 3\times 10^{13} \delta t_{-1} \Gamma_2^2~~~{\rm cm},
\end{equation}
where $c$ is the speed of light, $\delta t_{-1} = \delta t/ 0.1 $ s and $\Gamma_2 = \Gamma/100$.

The optical depth $\tau$ of the relativistic jet plasma is less than unity at radii greater than the photospheric radius $R_p$.
A conservative estimate for the minimum photospheric radius can be made by considering the electrons that accompany baryons in the jet.
By considering the scattering of photons by these electrons the optical depth can be estimated \citep[e.g.][]{2001ApJ...555..540L}.
At a radius $R$ the optical depth would be $\tau = \sigma_T E /\paren{4\pi R^2 m_p c^2 \Gamma}$, where $\sigma_T$ is the Thomson cross-section, $E=4\pi\epsilon$ is the isotropic equivalent blast energy, and $m_p$ is the mass of a proton.
The radius where $\tau = 1$ is the photospheric radius
\begin{equation}
R_p \simeq 6\times 10^{13} E_{52}^{1/2}\Gamma_2^{-1/2}~~~{\rm cm},
\end{equation}
where $E_{52}=E/10^{52}$ erg.

For a jet element with low-$\Gamma$ the initial dissipation happens well inside the photosphere;
due to the relativistic beaming effect the dynamics and emission for the element can be evaluated in the spherical model with isotropic equivalent energy $4\pi\epsilon$ and $\Gamma$.
The $\gamma$-rays of the prompt emission are injected into an optically thick medium and the photons can remain trapped.
The thermal energy of these trapped photons will be converted back to jet kinetic energy \citep{2001ApJ...551..934K, 2002ApJ...577..302K} and the prompt $\gamma$-rays from this jet region would be suppressed.
For an observer looking `on-axis' at such a region, all the prompt emission could be suppressed, resulting in a failed GRB \citep{2002MNRAS.332..945R}. 

For $\gamma$-rays injected below the photosphere, the energy density is adiabatically cooled until the photons de-couple at the photospheric radius.
The decoupling/emission time for these photons will be delayed from the dissipation or energy injection time $t_0$.
Dissipation occurs during the coasting phase of the jet where $\Gamma$ is constant and temperatures are sub-relativistic \citep{1993MNRAS.263..861P}.
As the energy density $e$ evolves as $e \propto R^{-8/3}$, and the injected luminosity evolves as $L_{\gamma} \Delta/c \propto eR^2\Delta \Gamma^2$, where $L_{\gamma}$ is the injected $\gamma$-ray luminosity and $\Delta$ is the shell width, the emitted $\gamma$-ray luminosity at the photosphere $L_{\gamma,p}$ will be
\begin{equation}
L_{\gamma,p} \simeq L_{\gamma} (R_p/R_\gamma)^{-2/3}~~~{\rm erg ~s^{-1}}.
\end{equation}
Additional to the adiabatic cooling, the prompt photons will be Compton downscattered and thermalized; 
the efficiency of the thermalization depends on the depth below the photosphere and therefore the optical depth \citep{2005ApJ...635..476P, 2007ApJ...666.1012T}.
The high energy spectrum will steepen and pair-production will determine a maximum spectral energy.
The low energy spectral slope will steepen due to Compton scatterings as the thermalization becomes more efficient.

A relativistic jet propagating into an ambient medium will decelerate when the swept-up mass is equivalent to $M_0/\Gamma$, where $M_0=4\pi\epsilon/\Gamma$ is the explosion rest mass.
A forward and reverse shock form and synchrotron radiation produces the observed afterglow of GRBs \citep[e.g.][]{1992MNRAS.258P..41R, 1997ApJ...476..232M, 1999ApJ...513..669K, 1999ApJ...520..641S}.
The deceleration radius is $R_{d} \propto l/\Gamma^{2/3}$ where $l$ is the Sedov length $l=\paren{3E/4\pi m_pc^2 n}^{1/3}$.
The observed deceleration time is then $t_{d} \propto E^{1/3}n^{-1/3}\Gamma^{-8/3}$.

A reverse shock will propagate through the ejecta from the central engine at the beginning of the decelerating blastwave phase.
The reverse shock contains energy comparable to the forward shock but due to a higher mass, the peak frequency is lower by a factor $\sim \Gamma^2$ \citep{2003ApJ...582L..75K}.
High polarization measurements in the afterglow of long GRBs suggests magnetized jets \citep{2009Natur.462..767S,2013Natur.504..119M}, these observations still support a baryonic jet rather than a Poynting flux dominated jet, although a strong magnetic field can suppress the reverse shock.
The reverse shock emission associated with short GRBs is rarely observed, either due to the early time of the peak, the typical frequency well below optical, or due to magnetic suppression.
We consider only the forward shock emission in this paper.

\subsection{Numerical Model}
\label{Model}
Jet parameters used throughout this paper are; bulk Lorentz factor $\Gamma=100$, ambient number density $n=0.1$ cm$^{-3}$, microphysical parameters $\varepsilon_B=0.01$, $\varepsilon_e=0.1$, $\gamma$-ray efficiency $\eta=0.1$, and minimum variability timescale $\delta t=0.1$ s;
the isotropic equivalent jet kinetic energy is $E_{\rm k}=E_{\rm iso}(1-\eta)$.
We have used an isotropic equivalent blast energy of $E_{\rm iso}=4\pi\epsilon_c=2\times10^{52}$ erg s$^{-1}$;
this value is taken from the peak of the $E_{\gamma,{\rm iso}}$ distribution in \cite{2015ApJ...815..102F}, and assuming our $\gamma$-ray efficiency.
The blast energy value is consistent with that found for jets from mergers by \cite{2017arXiv170504695S} and for the break-point in the luminosity function for short GRBs found by \cite{2015MNRAS.448.3026W}.

To estimate the observed intensity of the emission from a relativistic source at a generic viewing angle, we consider the Lorentz invariant quantity $I_\nu/\nu^3$, where $I_\nu$ is the specific intensity and $\nu$ the frequency \citep{1979rpa..book.....R}.
As $\nu=\delta~\nu'$, where $\delta=[\Gamma(1-\beta\cos\alpha)]^{-1}$ is the relativistic Doppler factor, $\Gamma=(1-\beta^2)^{-1/2}$ the bulk Lorentz factor and $\beta$ the velocity as a fraction of the speed of light, $\alpha$ the inclination to the line of sight of the bulk motion;
then $I_\nu=I'_{\nu'}\delta^3$ where primed quantities are in the co-moving frame.
By considering the observed on-axis emission, the specific flux to an off-axis observer will be a factor $a^3$ times the on-axis value, where $a=\delta(\alpha)/\delta(\alpha=0)<1$, i.e. $F_\nu(t,\alpha)=a^3F_{\nu/a}(at,\alpha=0)$ for a point source \citep{2002ApJ...570L..61G}.

We model the prompt and afterglow emission from compact stellar merger jets by dividing the jet structure into $N\times M$ segments defined using spherical co-ordinates;
the angle from the jet central axis is defined as $0<\theta_i<\theta_j$ and the rotation around the jet central axis as $0<\phi_k<2\pi$.
A segment has an opening angle of $\Delta\theta=\theta_j/N$ and an angular width $\Delta\phi=2\pi/M$.
The normal of each segment surface is $\theta_i$ from the central axis, where $\theta_i=(i-1/2)\Delta\theta$, $i$ is an integer in the range $1\leq i \leq N$.
Similarly, the rotation position is $\phi_k=(k-1/2)\Delta\phi$, $k$ is an integer in the range $1\leq k \leq M$.

A segment has a bulk Lorentz factor and energy consistent with the jet structure model used;
where for the jet structure models considered here, $\theta<<\theta_c$ (i.e. the segment next to the jet axis) is used to normalize the energy distribution.
Each segment has an energy per unit solid angle $\epsilon_{i,k}$ and a bulk Lorentz factor $\Gamma_{i,k}$.
The energy dissipated as $\gamma$-rays at the radius $R_\gamma\propto \Gamma_{i,k}^2$ is $L_{\gamma,i,k}\sim 4\pi\eta~\epsilon_{i,k}/t_{\rm in}$, where $t_{\rm in}$ is the energy injection timescale i.e. the pulse duration of $\gamma$-ray emission from a segment.
We assume that $t_{\rm in}\equiv\delta t$;
short GRBs often have multiple pulses, in such a case the duration of the prompt emission is longer than the variability timescale $t_{\rm in}>\delta t$, the choice of $t_{\rm in}=\delta t$ results in bright GRBs and it gives conservative estimates for the orphan afterglow rates.
The energy dissipated by each segment is then $t_{\rm in}L_{\gamma,i,k}\Omega_{i,k}/4\pi$.

{\bf Prompt emission:} The $E F_E \equiv \nu F_\nu$ spectrum for the injected photons is assumed to be a broken power-law that peaks at $E_p$ with a spectral index of 1.5 below the peak and -0.25 above the peak.
The spectral peak follows the $L_{\gamma}-E_p$ relation $E_{p,i,k}\sim300\paren{L_{\gamma,i,k}/10^{52}~{\rm erg}}^{2/5}$ keV \citep{2004ApJ...609..935Y,2009A&A...496..585G,2012ApJ...750...11Z},
where $L_{\gamma,i,k}$ is the isotropic equivalent $\gamma$-ray energy in the segment.
For each segment the optical depth at $R_\gamma$ is $\tau_{i,k}=(R_p/R_\gamma)^2$;
if $\tau_{i,k}>1$ then the photons will be coupled to the jet plasma until a radius $R_p$ when $\tau_{i,k}=1$ \citep{2011ApJ...737...68B, 2014ApJ...782....5H, 2016ApJ...829..112L}.
For cases where $\tau_{i,k}>1$ at $R_\gamma$, the photon energy will be adiabatically cooled as $L_{\gamma,i,k}\tau_{i,k}^{-1/3}$;
and the spectral peak energy will similarly reduce by a factor $\tau_{i,k}^{-1/3}$.
The condition for efficient thermalization is, $\tau_{i,k}\geq m_ec^2/k_BT_{BB}$ \citep{2005ApJ...635..476P,2007ApJ...666.1012T} where $m_e$ is the mass of an electron, $k_B$ is the Boltzmann constant, and $T_{BB}$ is the electron blackbody temperature $T_{BB}=(L_{\gamma,i,k}/4\pi R_\gamma^2\Gamma_{i,k}^2 c~a_c)^{1/4}$, here $a_c$ is the radiation constant.
If this condition is met then the spectral peak energy is given by $\sim3k_BT_{BB}$ and the spectrum is exponentially suppressed above this energy.
If thermalization is not efficient, then the maximum spectral energy is limited by pair-production;
a cut-off in spectral energy occurs at $511(\Gamma_{i,k}/\tau_{i,k})$ keV.

For each segment, the luminosity and timescales for an on-axis observer are determined using the fireball model.
The on-axis luminosity and time are corrected for the angle from the segment to the observers line-of-sight.
The emission time $t_e$ for each segment depends on the point at which the photons de-couple from the plasma.
For segments where $\tau_{i,k}\leq1$ this occurs at $t_0$;
for segments where $\tau_{i,k}>1$ then the emission is delayed so $t_e(\alpha=0)=t_0+(R_p-R_\gamma)/2\Gamma^2_{i,k}c$.
For an observer at $\theta_{\rm obs}$ and  $\phi_{\rm obs}$, the angle is $\alpha_{i,k}$.
The emission time for segments at an angle $\alpha_{i,k}$ is delayed, so $t_e(\alpha)=a^{-1}t_e(\alpha=0)$.
Since the dissipated energy is radiated over an area $D_L^2\Omega_{{\rm e},i,k}$, the on-axis flux is given by,
\begin{equation}
F_{\nu,i,k}(t,\alpha=0)=\frac{L_{\nu,i,k}}{4\pi D_L^2}\frac{\Omega_{i,k}}{\Omega_{{\rm e},i,k}},
\label{ON}
\end{equation}
where $\Omega_{{\rm e},i,k}={\rm max}[\Omega_{i,k},\Omega_{\Gamma,i,k}]$;
and $\Omega_{\Gamma,i,k}\paren{t_{i,k}}=2\pi(1-\cos1/\Gamma_{i,k})$ the beaming solid angle defined by the instantaneous segment bulk Lorentz factor.
Similarly, the frequency of the emission is lowered, and the duration will be longer, by the factor $a$.
The flux from each segment for an off-axis observer is given by,
\begin{equation}
F_{\nu,i,k}(t,\alpha_{i,k})=a^3~F_{\nu/a,i,k}(at,\alpha=0)~\cos\alpha_{i,k},
\label{flux}
\end{equation}
where $\cos\alpha_{i,k}$ is the correction for the emission area projection \citep{2003ApJ...592.1002S}.
The spectral peak is normalized as the value integrated between 1 keV and 10 MeV giving $L_{\gamma,i,k}$.
The prompt emission is then the sum of each segments emission in a time bin between $t_0$ and the maximum emission time $a^{-1}(t_e+t_{\rm in})$.
The burst is detected if the number of photons at the detector is $>0.2$ ph s$^{-1}$ cm$^{-2}$ in the {\it Swift} band, 15-150 keV \citep{2006ApJ...644..378B}.

{\bf Afterglow emission:} Jet energy that is not radiated away by the prompt emission drives a relativistic outflow into the interstellar medium.
The kinetic energy per unit solid angle of a jet segment is $\epsilon_{{\rm k},i,k}=\epsilon_{i,k}-t_{\rm in}L_{\gamma,i,k}/4\pi$.
We assume no sideways expansion so each jet segment evolves independentally \citep{2012ApJ...751..155V};
the lateral expansion of homogeneous and structured jets is discussed by \cite{2003ApJ...592.1002S}.
The value of $\Gamma_{i,k}$ is considered constant, $\Gamma_{0,i,k}$, before the deceleration radius $R_d$ and will evolve as $\Gamma_{0,i,k}(R_{i,k}/R_d)^{-3/2}$ with distance $R_{i,k}$ when $R_{i,k}>R_d$.
The on-axis flux from each segment at a given observer time $t$ can be evaluated by using the standard synchrotron shock model.
The on-axis characteristic frequency $\nu_m$ and cooling frequency $\nu_c$ are calculated in the same way as discussed in \cite{1998ApJ...497L..17S}.
The peak flux of the afterglow is obtained by considering the total number of electrons in the segment $N_e=nR^3\Omega_{i,k}/3$.
The total energy per unit time per unit frequency emitted by these electrons is proportional to $N_e \propto \Omega_{i,k}$ and is distributed over an area $D_L^2\Omega_{{\rm e},i,k}$ at a distance $D_L$ from the source.
Since the on-axis peak flux density $F_{\nu,{\rm max}}$ is proportional to $\Omega_{i,k}/\Omega_{{\rm e},i,k}$, we obtain the on-axis flux from a segment $F_{\nu,i,k}(t,\alpha=0)=F_{\nu}(t)~\Omega_{i,k}/\Omega_{{\rm e},i,k}$, where $F_\nu(t)$ indicates the flux from a blast wave with the isotropic energy $4\pi\epsilon_{{\rm k},i,k}$ \citep{1998ApJ...497L..17S,1999ApJ...519L..17S}.
For an off-axis observer the flux from a segment is given by equation \ref{flux};
the sum of flux from each segment at time $t$ gives a total afterglow light-curve.
Using this model the emission from a decelerating jet can be estimated at various observation angles.

\subsection{Homogeneous Jets: approximations}\label{H}

Here we give an approximation for the peak flux and peak time of an orphan afterglow from a homogeneous jet;
the estimates will be compared with the numerical results.
The afterglow emission from a decelerating relativistic collimated blastwave is beamed within the angle $\theta_j+1/\Gamma$.
For observers outside this angle the emission becomes much fainter as the inclination of the system increases.
Assuming slow cooling with $\nu_m<\nu<\nu_c$, and the Doppler correction for an off-axis observer, the observed peak flux is approximately,
\begin{equation}
F_p = C(p)~f(\theta_{\rm obs},\theta_j)~[\theta_{\rm obs}-\theta_j]^{2(1-p)}~\nu^{(1-p)/2}~E_{\rm k}~n^{(1+p)/4}~\varepsilon_B^{(1+p)/4}~\varepsilon_e^{p-1}~D^{-2}~~~{\rm erg ~s^{-1}~cm^{-2}~Hz^{-1}},
\end{equation}
where $C(p)$ is a constant that depends on the particle index $p$ and all the relevant physical parameters\footnote{$C(p)=(32\pi)^{(1+p)/4}~(12\pi)^{-1}~(2\pi)^{(1-p)/2}~m_p^{(5p-7)/4}~m_e^{(5-3p)/2}~q_e^{(p-3)/2}~c~\sigma_T~[(2p-4)/(7-p)]^{p-1}~[(7-p)/(5+p)]^{(5+p)/2}$, where $m_e$ is the electron mass, $c$ the speed of light, $\sigma_T$ the Thomson cross-section, $q_e$ an electron charge, $m_p$ the proton mass}, $f(\theta_{\rm obs},\theta_j)$ accounts for the jet opening angle $\theta_j$, viewing angle $\theta_{\rm obs}$ and the relativistic beaming, and $\nu$ is the observed frequency.
The factor $f(\theta_{\rm obs},\theta_j)$ is,
\begin{equation}
f(\theta_{\rm obs},\theta_j)=\cos{(\theta_{\rm obs}-\theta_j)}\left[\frac{1-\cos\theta_j}{1-\cos\left([(7-p)/(2p-2)]^{1/2}~\theta_{\rm obs}\right)}\right],
\label{norm}
\end{equation}
where $\cos(\theta_{\rm obs}-\theta_j)$ corrects for the surface area projection, and the second term accounts for the emission solid-angle.

For $p=2.5$, the peak flux is,
\begin{equation}
\begin{split}
F_p \sim 2\times10^{-3}~f(\theta_{\rm obs},\theta_j)~[\theta_{\rm obs}-\theta_j]^{-3}~\nu_{14}^{-3/4}~E_{52}~n_{-1}^{7/8}~\varepsilon_{B,-2}^{7/8}~\varepsilon_{e,-1}^{3/2}~D_{200\rm{Mpc}}^{-2}~~~{\rm mJy},
\end{split}
\label{oa}
\end{equation}
where we use the convention $N_x=N/10^x$.
Angles are in radians, frequency is in Hz, $E$ is the isotropic jet kinetic energy $E_{\rm k}$ in erg, ambient number density $n$ in cm$^{-3}$, and the distance is normalized to 200 Mpc.

The peak flux occurs at a time given by,
\begin{equation}
t_p \sim 195~ \left[\frac{(5+p)(7-p)^{1/3}}{(p-1)^{4/3}}\right]\left[\theta_{\rm obs}-\theta_j\right]^{8/3}~n_{-1}^{-1/3}~E_{52}^{1/3}~~~{\rm days}.
\label{oat}
\end{equation}
The expressions in equation \ref{oa} and \ref{oat} give an approximation for the peak flux and time from an off-axis orphan afterglow to a relativistic jet with homogeneous structure in a uniform density ambient medium.

\subsection{Monte Carlo Results}
\label{MC}

\begin{figure*}
\includegraphics[scale=0.8]{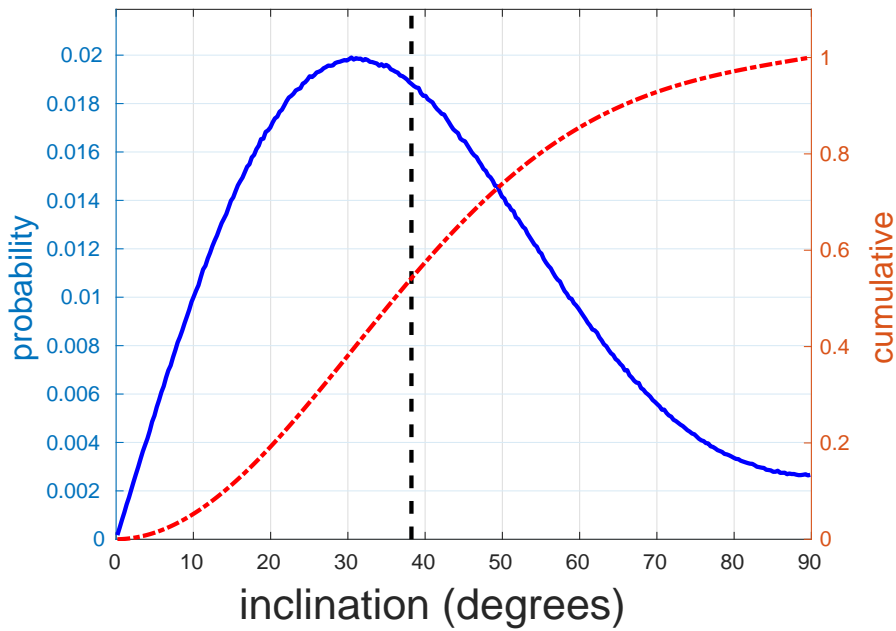}
\caption{By considering the GW strain from a merger as a function of inclination the distribution of system inclinations can be determined. For all GW detected mergers at a fraction of the maximum detectable luminosity distance the probabilty of a system being inclined at given angle is shown with the blue solid line. The mean system inclination for this distribution is the dashed black line. The red dashed-dotted line is the cumulative distribution.}
\label{fig.GWi}
\end{figure*} 

Given a GW detection from a NS-NS or BH-NS merger, the fraction of events that have detectable EM counterparts from the relativistic jet depends on the jet structure and opening angle.
Using a Monte Carlo method we estimate the fraction of merger jets, with a given jet structure, that result in EM counterparts brighter than ${\it r}$-band magnitude $m_r\leq21$.
A population of $10^5$ mergers within the face-on detection limit for a NS-NS merger by advanced LIGO $\sim300$ Mpc is generated.
The luminosity distance $D_L$ to a merger is randomly determined using the redshift distribution for non-collapsar short GRBs found by \cite{2015MNRAS.448.3026W}.
The inclination $i$ follows a random isotropic distibution.
By considering that GW signals are stronger along the system rotation axis for binary mergers with a random orientation, the average inclination for a distribution of GW detected mergers can be determined.
Mergers with a GW strain $h\propto (h_+^2+h_\times^2)^{1/2}/D_L$, where $h_+\propto1+\cos^2i$ and $h_\times\propto 2\cos i$, are GW detected if $h > h_c$ the limiting detectable strain \citep[e.g.][]{1993ApJ...417L..17K, 2016ApJ...829..112L};
for a more detailed investigation of the detectable gravitational waves from compact binary mergers see \cite{2003ApJ...589..861K,2010ApJ...725..496N,2011CQGra..28l5023S}.
The distribution of merger inclinations is shown in figure \ref{fig.GWi};
the peak of the probability distribution is $i\sim 31^\circ$, and the mean $\langle i \rangle \sim 38^\circ$.
The blue solid line is the probability of a merger with a given inclination;
the red dash-dotted line is the probability that a merger will have an inclination equal or less than a given value.

\begin{figure*}
\includegraphics[width=\columnwidth]{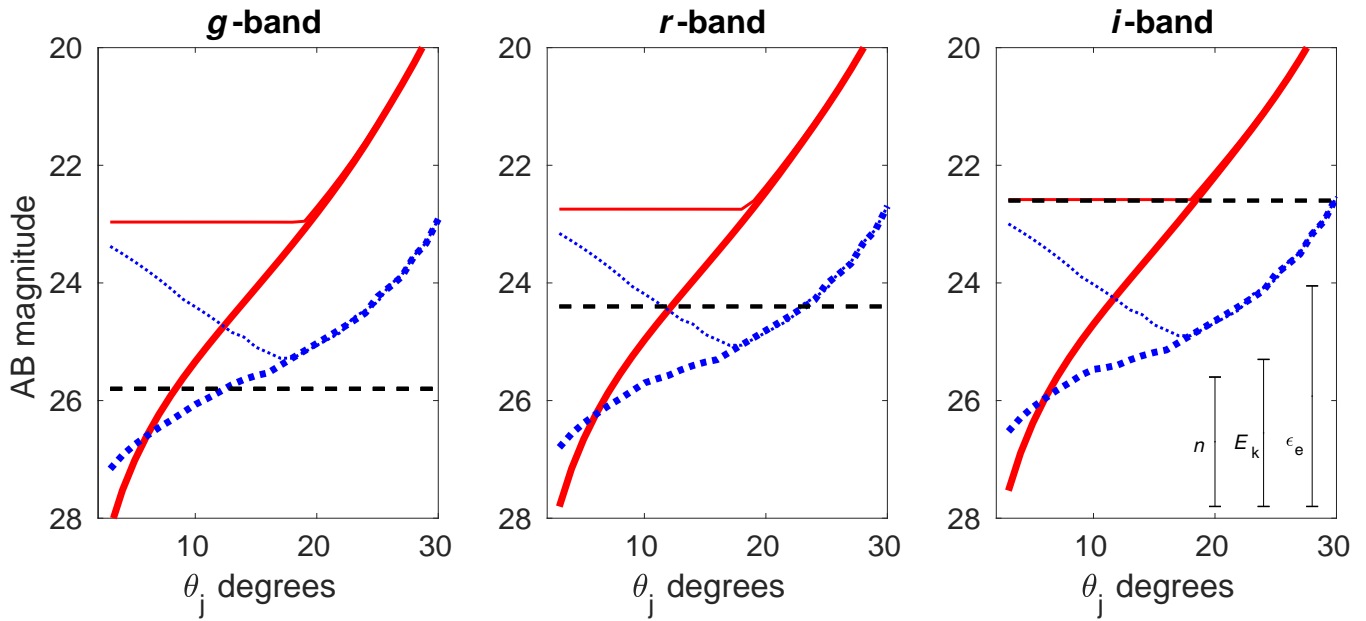}
\caption{The peak magnitude for the off-axis afterglow at 38$^\circ$ from a homogeneous merger jet with opening angle $\theta_j$ at 200 Mpc. Red thick line indicates a jet with constant isotropic equivalent energy, $E_{\rm iso}=2\times10^{52}$ erg, and the blue thick dashed line indicates a jet with constant geometrically corrected jet energy (normalized to a $\theta_j=6^\circ$ jet with $E_{\rm iso}=2\times10^{52}$ erg). The thin lines indicate a two component jet where $\theta_j$ defines the core angle ($\theta_c$ in equation \ref{eq_4}) and the wider component extends to 30$^\circ$ (equivalent to $\theta_j$ in equation \ref{eq_4}). The energy and Lorentz factor of the wide component are fixed at 5\% the core values. All jets have a core Lorentz factor of $\Gamma=100$ and are in an ambient medium with a particle density of $0.1$ cm$^{-3}$. The full size of the errorbars in the righthand panel indicate the magnitude of change in peak flux for a one order of magnitude change in the respective parameter (note that $n$ is degenerate with $\varepsilon_B$). The black dashed horizontal lines indicate the peak macronova emission for a NS-NS merger; assuming isotropic emission from a soft equation-of-state model e.g. Tanaka et al. (2014) at 200 Mpc}
\label{fig.H10}
\end{figure*}

The peak magnitude for an observer at the mean GW detection inclination angle of $\sim38^\circ$ from a homogeneous jet depends on the half-opening angle of the jet.
By considering a homogeneous jet with a constant isotropic equivalent blast energy, or a constant geometrically corrected jet energy, the peak magnitude of the orphan afterglow for an observer at 200 Mpc and 38$^\circ$ can be estimated.
Using the isotropic equivalent energy $E_{\rm iso}=2\times10^{52}$ erg, or the geometrically corrected energy $E=E_{\rm iso}\Omega/2\pi=3\times10^{50}$ erg, giving $E_{\rm iso}$ for a $\theta_j=10^\circ$, the peak magnitude for jet half-opening angles $2^\circ\leq\theta_j\leq30^\circ$ are shown in figure \ref{fig.H10}.
The thick red line is for constant $E_{\rm iso}$, and the thick blue dotted line for constant geometrically corrected jet energy.
Three optical bands are shown, {\it g}-, {\it r}-, and {\it i}-band and the equivalent peak macronova flux, black dashed line, for a NS-NS merger \citep{2014ApJ...780...31T}.
BH-NS mergers would result in brighter macronova, $\sim 23.8,~23.2,~22.8$ respectively, although the ejecta in these cases is not isotropic.
The macronova estimates should be considered as upper limits{, for the adopted model,} as the peak flux depends on the inclination where the brightest emission coincides with the polar axis (the jet axis) \citep{2016AdAst2016E...8T,2017arXiv170507084W}{;
however, macronova may be brighter than the adopted model i.e. \cite{2016NatCo...712898J}}.
The frequency dependence for the afterglow flux is shallower than that of a macronova which peaks sharply in the red to radio with a thermal spectrum and exponential decay at higher frequencies.
The non-thermal spectrum of a GRB afterglow, where the higher frequency is typically $F_\nu \propto \nu^{-(p-1)/2}$ or $F_\nu \propto \nu^{-p/2}$ where $p\sim2.5$, ensures that for an off-axis observer the afterglow is at a similar amplitude in a range of detection bands.

In figure \ref{fig.H10} we see that the peak flux for an orphan afterglow viewed at 38$^\circ$ is brighter for homogeneous jets with wider jet half-opening angles.
The point at which the peak flux for constant isotropic equivalent blast energy and constant geometrically corrected jet energy are equal indicates the normalization angle.
For jets normalized to this value with narrower half-opening angles, the peak afterglows are brighter than the equivalent constant isotropic blast energy case;
this is due to the jet having a higher energy density in these cases, for jets wider than this normalization, a reduction in jet energy density is apparent.
The shape of the curve is dominated by the effective angle to the jet for wide $\theta_j$ i.e. $(\theta_{\rm obs}-\theta_j)^{-3}$ equation \ref{oa};
and for narrower $\theta_j$, by the fill factor i.e. the second part of the expression in equation \ref{norm}.
{For a jet with a given opening angle, inclination, distance, and observation frequency the peak orphan afterglow flux is $F_p\propto E_{\rm k}n^{7/8}\varepsilon_B^{7/8}\varepsilon_e^{3/2}$; the degeneracy in $\varepsilon_B$ and $n$ can make determination of these parameters difficult;
the change in peak flux for a one order of magnitude change in any of these parameters is indicated by the length of the errorbars in the third panel.
Short GRBs often occur in low-density environments, a reduction in $n$ by an order of magnitude would result in a peak that is $\Delta m_{AB}\sim2.2$ dimmer.}

Within figure \ref{fig.H10} the peak flux for the orphan afterglow of a two-component jet is shown as a thin red and a thin blue dotted line.
In each case the wider jet structure extends to $30^\circ$ (equivalent to $\theta_j$ in equation \ref{eq_4}) with energy and Lorentz factor at 5\% the value for the core region, defined by the x-axis in the figure.
For the thin red line the jet has an isotropic equivalent blast energy for an on-axis observer $\theta_{\rm obs}<\theta_c$ of $2\times10^{52}$ erg;
the thin dotted blue line has a constant geometrically corrected jet energy normalized to a homogeneous jet with an opening angle of $6^\circ$. 
As the two-component jet always has a wide sheath that extends to 30$^\circ$, beyond the core angle defined by $\theta_j$ on the plot x-axis, the peak flux for jets with a core narrower than $\sim20^\circ$ is constant and approaches the homogeneous jet case for half-opening angles wider than this.
By considering equation \ref{oa} for two homogeneous jets, one with fixed energy and undefined $\theta_j$ and the second with $\theta_j=30^\circ$ and 5$\%$ the energy of the first, the $\theta_j$ for the more energetic jet that results in the same peak orphan afterglow for an observer at $\theta_{\rm obs}$ is $\theta_j\sim\theta_{\rm obs}-20^{1/3}(\theta_{\rm obs}-30)$ degrees.
The wide sheath with 5\% the core energy and Lorentz factor is the dominant contributer to the off-axis emission for jets with a core $\la20^\circ$.
Where the jet energy is fixed at the geometrically corrected value for a 6$^\circ$ homogeneous jet, the reduction in the energy content of the wider component as the core width is increased leads to a dimmer afterglow.
When the off-axis emission from the jet core becomes brighter than the off-axis emission from the sheath, the peak off-axis flux follows the homogeneous jet.
Two-component jets are described in \S\ref{structure} and their afterglows discussed below.

The Monte Carlo distribution of mergers for each structure model {have identical values of the core opening angle $\theta_c=6^\circ$.}
Hydrodynamic simulations indicate a range of jet core half-opening angles that are dependent on the initial conditions, $3^\circ\la\theta_c\la13^\circ$ \citep{2014ApJ...784L..28N}.
The core value is significantly wider than the core values used in other structured jet models \citep[e.g.][]{2002MNRAS.332..945R,2003ApJ...592.1002S}.
The two-component jet has $\epsilon_s$ and $\Gamma_s$ at 5$\%$ the core values, while the power-law jet has an index $k=2$ for $\theta>\theta_c$.
The effect of jet structure on the observed jet-break is discussed below.
For the extended structure the minimum $\Gamma$ is 2, and the maximum half-opening angle is 25$^\circ$, and all other parameters are as previously used. 

\begin{figure*}
\includegraphics[width=\columnwidth]{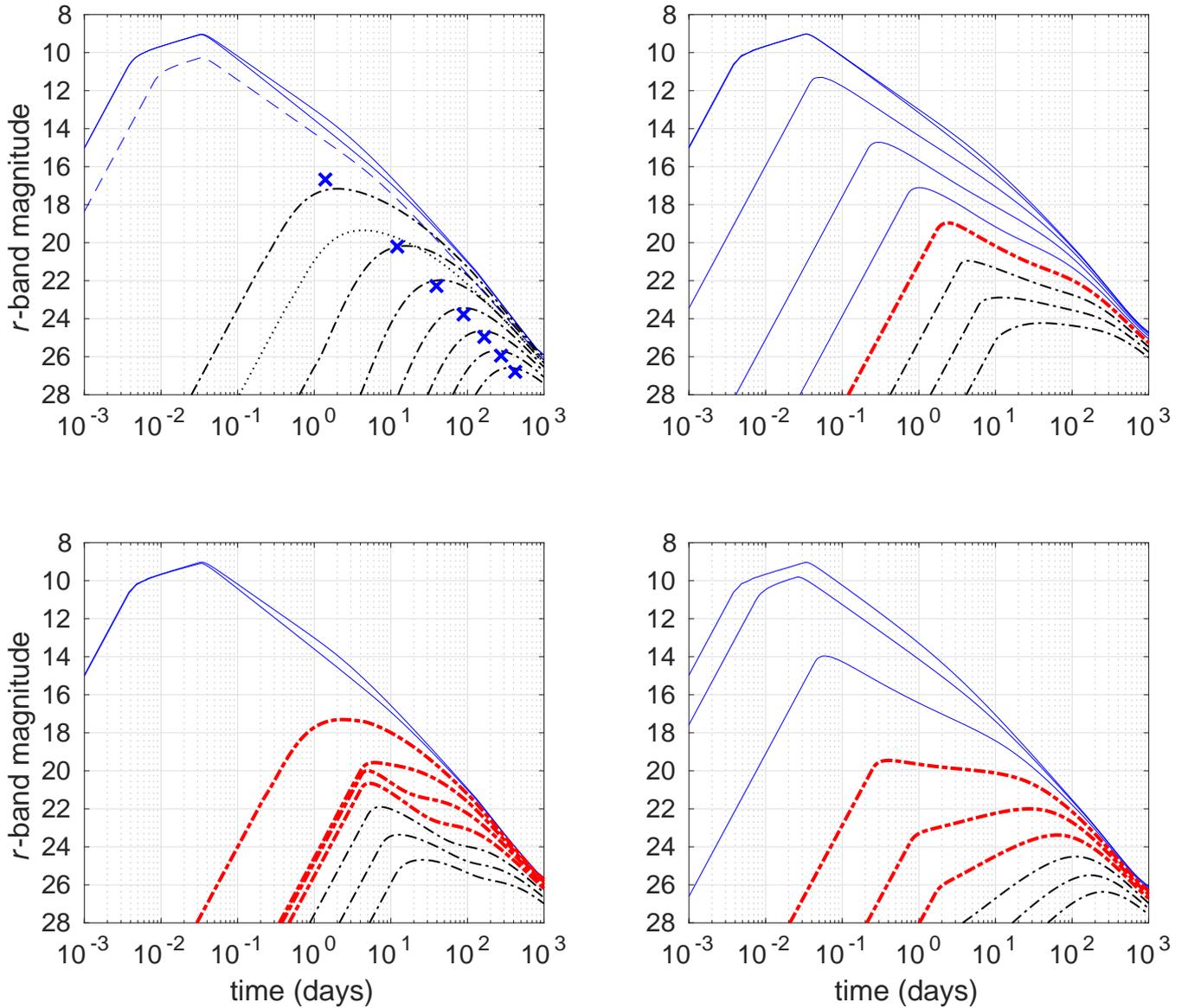}
\caption{Afterlow {\it r}-band lightcurves for jets at 200 Mpc. Lightcurves are plotted for an observer at 5$^\circ$ increments in the range $0^\circ \leq \theta_{\rm obs} \leq 40^\circ$. The model values used in each case are: (top left) $\theta_c=\theta_j=6^\circ$ for the homogeneous jet; (bottom left) $\theta_c=6^\circ$ for the two-component jet where the second component extends to $\theta_j=25^\circ$ with 5$\%$ of the core energy and Lorentz factor; (top right) $\theta_c=6^\circ$ for the power-law jet with an index $k=2$ for $\theta_c<\theta\leq25^\circ$; and (bottom right) $\theta_c=6^\circ$ for the Gaussian jet with a maximum $\theta_j=25^\circ$. Jets have an isotropic equivalent blast energy of $2\times 10^{52}$ erg, a bulk Lorentz factor $\Gamma=100$, and an ambient medium density of $n=0.1$. Blue lines indicate the afterglow of a GRB; red dashed lines indicate an on-axis orphan afterglow i.e. within the wider jet opening angle but with suppressed prompt emission; black dashed-dotted lines indicate an off-axis orphan afterglow. The blue dashed and black dotted lines in the top left panel indicate the afterglow for an observer at 0$^\circ$ and 10$\circ$ respectively where the ambient medium has a particle density $n=0.01$ cm$^{-3}$; the change in magnitude for an `on-axis' observer is $\Delta m_{r}\sim1.2$, and for an `off-axis' observer $\Delta m_{r}\sim2.2$ for each order of magnitude change in the $n$ parameter.}
\label{fig.10LC}
\end{figure*}

Examples of the afterglow light-curves for each model from a jet at 200 Mpc and viewed at inclinations from 0$^\circ$ to $40^\circ$, in $5^\circ$ intervals, are shown in figure \ref{fig.10LC};
each jet structure has $120\times120$ segments.
The lightcurve produced using $N=M=120$ in the model is identical for values of $N,~M>120$;
where $N,~M<120$ the peak flux and time for afterglows are consistently reproduced although the shape of the early afterglow before the peak is inaccurate.
Off-axis lightcurve shape is generally unaffected by the reasonable choice of segment number.
The blue lines indicate the afterglow for a {\it Swift} detectable GRB, $\theta_{\rm obs}\leq10^\circ$;
the red dashed lines indicate the afterglow for a jet viewed within the half-opening angle but without a {\it Swift} detectable GRB, a failed-GRB, $\theta_{\rm obs}\leq\theta_j$;
the black dash-dotted lines indicate an off-axis orphan afterglow, $\theta_{\rm obs}>\theta_j$.
For the homogeneous jet, the analytic peak magnitude and time from equations \ref{oa} and \ref{oat} are shown as blue crosses;
the analytic expressions overestimate the peak flux, and underestimate the peak time when $\theta_{\rm obs} \la 3\theta_j$.
{Additional lightcurves are shown in the top-left panel for an observer at 0$^\circ$ and 10$^\circ$, blue dashed and black dotted lines respectively.
Here the ambient number density is lower by a factor 10;
for an on-axis observed afterglow, this parameter change results in a peak flux that is $\sim1.2$ magnitudes fainter and for off-axis observed afterglow the peak flux is $\sim2.2$ magnitudes fainter.
A similar change in magnitude, $1.2 \la \Delta m_{r} \la 2.2$, is observed for all lightcurves where the ambient density is lower by a factor 10.}

The light-curves in figure \ref{fig.10LC} have afterglows which in each case are similar for an observer on the jet axis i.e. the deceleration time, peak flux, and peak time.
The jet has a soft break that is determined by either the difference between the obseravtion angle and the jet half-opening angle for a homogeneous jet, or the core angle for a jet with structure.
A second break may be observed at later times, this is associated with the opening angle of the extended structure.
A GRB afterglow for a homogeneous jet observed at the jet edge $\theta_j$ is half as bright and has a jet-break determined by the width of the jet $\sim 2\theta_j$;
for the other structures the afterglow characteristics depend on the local jet energetics $\epsilon$ and $\Gamma$ parameters.

Light-curves for the jet structure models tested show that, where no sideways expansion is assumed and the jet-break is caused by the increase in the beaming angle beyond the jet edge, that the break seen in short GRB afterglows depends on the inclination.
We expect a sharp break at very late times when the outflow becomes Newtonian, this is not included in our model.
\cite*{2015ApJ...815..102F} list four short GRBs with measured half-opening angles $3^\circ \la \theta_j \la 8^\circ$, and a further seven with lower limits;
the narrowest of these lower limits is $\ga 4^\circ$, and the widest $\ga 25^\circ$.
The average $\theta_j$ for short GRBs can be inferred by assuming a maximum jet half-opening angle;
$\bar{\theta_j}=16^{\circ~+11}_{~~-10}$ for $\theta_{\rm max}=30^\circ$, and $\bar{\theta_j}=33^{\circ~+38}_{~~-27}$ at the limit $\theta_{\rm max}=90^\circ$;
{alternatively, \cite{2016A&A...594A..84G} found the short GRB population to be consistent with a jet opening angle of $3\leq\theta_j\leq6^\circ$.
We use a $\theta_j=6^\circ$ for homogeneous jets, consistent with both estimates, and fix this as the core angle for jets with extended structure.}
In these examples the jet half-opening angle was inferred using $\Gamma^{-1}(t_b)\equiv \theta_j$.
If the observed jet-break time $t_b$ depends on inclination, as in our model for GRB afterglows, the break time cannot limit the full extent of jet structure.
By assuming a range of jet parameters, the range of jet-break times can be reproduced by our model.

Additional features in the afterglow light-curves for jets with extended structure appear at wider angles.
For our parameters, these appear where the prompt emission is supressed and the afterglow would be from a failed-GRB.
Afterglows for the two-component model at angles $\theta_{\rm obs}> \theta_c$ have an early peak flux and time determined by the local jet energy $\epsilon$ and $\Gamma$ respectively;
a late bump is due to emission contribution from the bright core, the time of the bump is determined by the inclination, with higher inclinations resulting in a later bump time.
A similar feature can be seen in the power-law structured jet but as the energetics and Lorentz factor for the wider component are not uniform with angle, the early peak flux and time are unique.
The afterglow for the Gaussian structured jet at comparable angles is dominated by the bright core emission at late times.
For orphan afterglows in each structured jet case, the early rise time and peak are due to the contribution from the wide extended structure;
a more energetic wide component leads to a brighter and more pronounced peak, while for a less energetic wide component, the orphan afterglow is dominated by the core emission at later times.
As the observation angle increases, the contribution from the various components becomes indistinguishable, here we only show orphan afterglows until an observation angle of 40$^\circ$.

In all cases we have assumed uniform baryon loading;
if the baryon loading is more efficient towards the edge of a jet then $\epsilon$ and $\Gamma$ will not have the same distribution.
If the structure in a jet is due to baryon loading only, then the energy will be uniform;
the afterglow for the various viewing angles will be brighter than the equivalent shown here as the peak flux depands on the energy.
The peak time for the afterglow will be later for lower-$\Gamma$ components; the prompt emission will be similarly suppressed.

\section{Peak Flux/Time}
\label{disc}

\begin{figure*}
\includegraphics[scale=0.6]{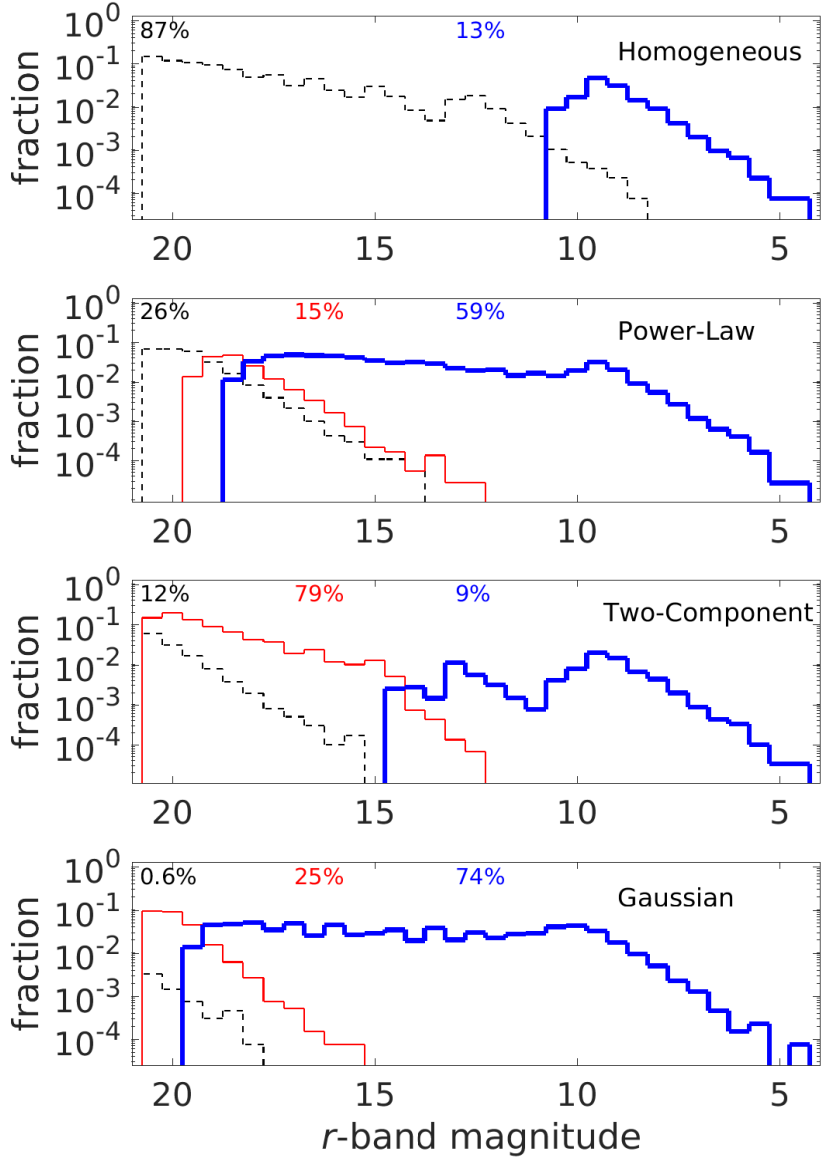}
\caption{Peak magnitude for the afterglow brighter than 21 for a population of 10$^5$ GW detected mergers; the percentage of the total detected merger population for each type is: homogeneous jets $13.6\%$; power-law structured $36.9\%$; two-component $30.0\%$; and Gaussian $13.3\%$. Blue thick line histogram is a GRB afterglow; red thin line histogram is an on-axis orphan (failed GRB) afterglow $\theta_{\rm obs}<\theta_j$; black dashed line is an off-axis orphan afterglow $\theta_{\rm obs}>\theta_j$. Percentages are the fraction of events brighter than magnitude 21 in each case.}
\label{fig.10MC}
\end{figure*}

The 10$^5$ Monte Carlo distribution has a randomly determined inclination and distance given a GW detection, the same distribution is used with each jet structure, the afterglow from each jet structure model is evaluated at one degree intervals $0^\circ\leq\theta_{\rm obs}\leq90^\circ$;
for efficiency, the model uses $N=25$ and $M=100$ ensuring jet structure is resolved.
The peak magnitude for the light-curve that corresponds to the jet structure at the randomly determined inclination is then selected and scaled for the distance.
A histogram of the peak magnitude for jet EM counterparts brighter than magnitude 21, for GW detected mergers $\la300$ Mpc is shown in figure \ref{fig.10MC};
the thick blue line is a GRB afterglow, the thin red line is a failed-GRB orphan afterglow, the black dashed line is an off-axis orphan afterglow.
The fraction of each jet counterpart type i.e. GRB afterglow, failed GRB afterglow, orphan afterglow, of the total number of $m\leq21$ events are shown.

In figure \ref{fig.10MC} the peak of the distribution for GRB afterglows is that for a face-on NS-NS merger at the maximum detection distance $\sim 300$ Mpc.
The structured jets have an extended distribution to fainter magnitudes when compared with the homogeneous jets, this is due to the lower energetics for observers $\theta_c<\theta_{\rm obs}$.
For the failed-GRB orphan afterglows from jets with structure, the distribution for power-law structured and Gaussian structured jets has a wide plateau for the peak magnitudes due to the non-uniform energetics of the wider jet component.
The two-component jet structure has a uniform energy distribution in the wide component,
this gives a single sharp peak to the failed-GRB orphan afterglows.

From the Monte Carlo the fraction of afterglow counterparts brighter than magnitude 21 depends on the jet structure model.
For jets with extended structure to the limit of 25$^\circ$, we show that compared to a population of homogeneous jets with $\theta_j=6^\circ$ the fraction of bright jet counterparts is higher for two-component jets (equation \ref{eq_4}) and power-law structured-jets (equation \ref{eq_2}).
{GRB producing jets result in bright afterglows, with peak {\it r}-band magnitude $20\ga m_r \ga 5$.}
Orphan afterglows brighter than magnitude 21, both from failed-GRBs and off-axis observations, are produced in {$\sim 12\%$ of cases for homogeneous jets;
$\sim 27\%$ for two-component jets;
$\sim 15\%$ of cases for power-law structured-jets;
and $\sim 3.4\%$ for Gaussian jets.}
The brightest of these counterparts is $m_r\ga8$.
The peak brightness depends on the jet kinetic energy and the fraction of events depends on the jet opening angle.
For mergers that are close by, the prompt photon flux at angles $>\theta_c$ can be above the detection threshold;
for two-component jets, where the $\epsilon$ distribution is generally flat in this region, a noticable fraction of the counterparts will accompany faint GRBs.
This can be seen by three peaks in the flux distribution for GRB afterglows.

The total fraction of EM counterparts brighter than magnitude 21 from the jet of GW detected mergers depends on the jet structure:
for homogeneous jets we find $\sim13.6\%$;
for two-component jets $\sim30\%$;
for structured jets the fraction is $\sim37\%$;
and Gaussian jets $\sim13\%$.
The fractions for an isotropic distribution to a distance of $\sim$200 Mpc, the maximum for edge on NS-NS GW detection, are $\sim 4.5\%,~11.8\%,~13.5\%,~{\rm and}~4.1\%$ respectively (homogeneous, two-component, power-law, and Gaussian);
here GRB afterglows account for $\sim 4.4\%,~3.4\%,~43.7\%,~{\rm and}~53.7\%$ of the $m_r\leq21$ counterpart fraction.
In all cases we consider the same structure parameters.
The fraction of events brighter than magnitude 10, in each case, are dominated by GRB afterglows.

\begin{figure*}
\includegraphics[scale=0.6]{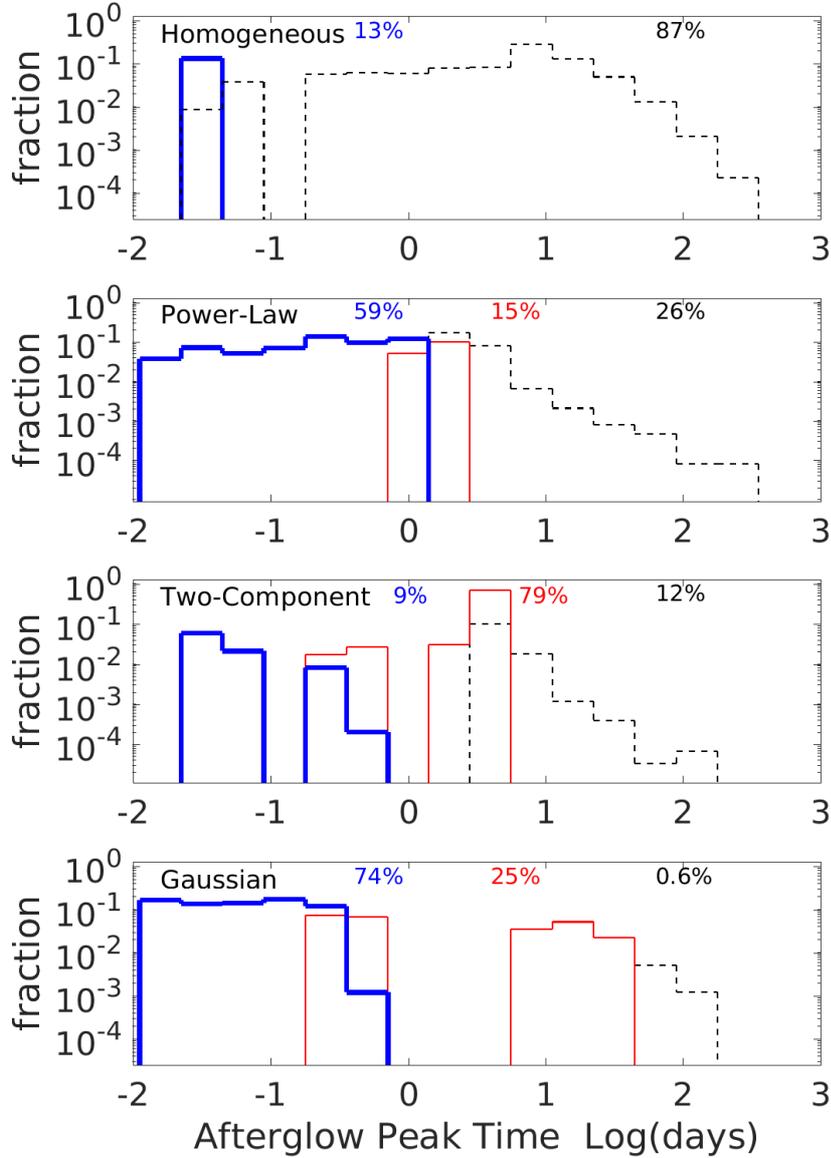}
\caption{Peak time for the afterglows brighter than 21 for a population of 10$^5$ GW detected mergers. Blue thick line histogram is a GRB afterglow; red thin line histogram is an on-axis orphan (failed GRB) afterglow $\theta_{\rm obs}<\theta_j$; black dashed line is an off-axis orphan afterglow $\theta_{\rm obs}>\theta_j$. Percentages are the fraction of events brighter than magnitude 21 in each case.}
\label{fig.10TP}
\end{figure*}

A corresponding histogram showing the peak time for each of the counterpart distributions is shown in figure \ref{fig.10TP}.
The colour and line style are the same as figure \ref{fig.10MC}.
The peak time distribution shows that for structured jets the GRB afterglows have a broader range of peak times than the homogeneous jet case.
This is due to the non-uniform distribution of Lorentz factor {for GRB producing jet components $>\theta_c$}.

The jet counterparts $m_r\leq21$, to GW detected mergers, typically peak at $t_p\la100$ days.
The brightest counterparts peak very early $0.01\la t_p\la0.1$ days;
orphan afterglows for a homogeneous jet peak typically at $t_p\sim10$ days;
failed-GRB and off-axis orphan afterglows typically peak at $t_p\sim1$ day for power-law structured-jets and two-component jets;
and Gaussian jets exhibit a bimodal distribution, due to the wide low-$\Gamma$ extended jet structure, that peaks at $t_p\sim0.25$ days and $t_p\sim20$ days.
The bimodal feature in the GRB afterglow distribution for two-component jets is due to the stepped boundary between the spine and sheath;
detectable GRBs are produced outside of the core region $\theta_c=6^\circ$, these GRBs near the core edge have significantly lower $\Gamma$ than those observed within the core angle;
the second split in peak times for the two-component jets is due to the dominence of the off-axis core emission over the on-axis sheath emission, where on-axis emission will peak earlier.
The apparent bi-modality of the bright orphan afterglows for a homogeneous jet is a result of the sharp jet edge and uniformity of $\theta_j$ for the population as well as the numerical precision for changes of inclination $<1^\circ$;
the bi-modality would vanish for a population of jets with a distribution of $\theta_j$ or higher numerical resolution.

\section{Conclusions}
\label{conc}

For jets from compact-stellar-mergers with a homogeneous structure we have shown that wide opening angles $\theta_j\ga10^\circ$ result in optical orphan afterglows, when viewed at the average GW detected merger inclination of $\sim38^\circ$, that are brighter than the estimates for the equivalent peak flux from macronovae{;
note that this depends on the ambient density and jet energetics}.
We show that where jets have an extended structure to a limit of $\theta_j=25^\circ$, similar to the limit predicted by numerical simulations, the fraction of EM counterparts brighter than magnitude 21 can be {2-3 times} that from a narrower homogeneous jet population.
GW triggered searches for EM counterparts could reveal a hidden population of failed-GRB orphan afterglows associated with wider jet structure where the low energetics and Lorentz factor could suppress the prompt $\gamma$-rays;
we show lightcurve features in orphan afterglows that could indicate the presence of extended jet structure. 
Jet EM counterparts to GW detected NS-NS or BH-NS mergers will reveal the jet structure, Lorentz-distribution, and opening angle for short GRB jets.

We assumed a jet central axis observed isotropic blast energy of $2\times10^{52}$ erg s$^{-1}$.
A jet with a higher blast energy will result in an afterglow with a brighter peak magnitude.
The various structured jet models naturally predict a range in observed total energetics that have a maximum at $2\times10^{52}$ erg s$^{-1}$.
The observed energetics of a jet, inferred from the prompt fluence and the peak of the afterglow, will appear lower for GRB afterglows seen at the jet edge for homogeneous jets or outside the jet core for jets with a variable structure.
Jets viewed at inclinations where most of the prompt emission is supressed may appear as X-ray flashes or low-luminosity GRBs;
in both cases the afterglow will appear dimmer and peak at later times than for the on-axis afterglow.
For jets observed at inclinations comparable to the point where $\gamma$-rays become suppressed, the duration of the prompt emission will be longer due to the delayed emission of the prompt photons from the low-$\Gamma$ segments;
the spectra will have a strong thermal contribution.
The longer duration of such a GRB could result in misclassification as $T_{90}\ga2$ s.

The rate of NS-NS mergers within the advanced LIGO detection volume is not known but values range from $0.2-200$ yr$^{-1}$ \citep[e.g.][]{2013PhRvD..88f2001A,2016LRR....19....1A}.
\cite{2012ApJ...746...48M} made an estimate for the {\it Swift} detected short GRB with redshift rate within this volume for NS-NS, 0.03 yr$^{-1}$;
{similarly, \cite{2012MNRAS.425.2668C,2013ApJ...767..140P,2014MNRAS.437..649S} found a consistent rate for GW-GRBs within the aLIGO volume, although the limits vary, and \cite{2015ApJ...815..102F} a merger-rate of $8^{+47}_{-5}$ yr$^{-1}$ which results in the same rate for {\it Swift}/BAT short GRBs from jets with an opening angle of 16$^\circ$ and the {\it Swift}/BAT field of view.}
{The {\it Swift}/BAT field of view is $\sim 1.4$ sr, therefore the all-sky rate of short GRBs within the NS-NS detection volume is $\sim0.27$ yr$^{-1}$;
or by assuming that all {\it Swift}/BAT GRBs have the same redshift distribution, the rate becomes $\sim1.1$ yr$^{-1}$ as only $1/4$ of {\it Swift}/BAT short GRBs have a measured redshift.
For each of our jet models we find the fraction that have peak afterglows brighter than $m_r\leq21$, of this fraction we find the percentage that are associated with GRBs.
If the all-sky rate of short GRBs within the NS-NS LIGO detection volume is $0.27\leq R_{SGRB}\leq1.1$ yr$^{-1}$ then the merger rate will be:
for homogeneous jets where 13\% of the GW detected population are $\leq21$, and $\sim13\%$ of these are GRB afterglows, the fraction of the total population that produces GRBs is $\sim1.7\%$ giving a GW detectable merger rate of $15.9\leq R_{NS-NS} \leq 63.5$ yr$^{-1}$;
for two-component jets, the fractions of the population that results in a detected GRB is $\sim2.7\%$, the merger rate is then $10\leq R_{NS-NS}\leq40$ yr$^{-1}$;
for power-law jets, GRB fraction is 22\%, and the merger rate is $1.2\leq N_{NS-NS}\leq 4.9$ yr$^{-1}$;
for Gaussian jets, the GRB fraction is 9.6\%, and the merger rate $2.8\leq R_{NS-NS}\leq 11.3$ yr$^{-1}$.}

{If we consider the number of potential counterparts that are brighter than $m_r\leq21$ for each of these models with our parameters, we find that homogeneous jets will result in $\sim2-8$ yr$^{-1}$, two-component jets will result in $\sim3-12$ yr$^{-1}$, power-law jets $\sim0.4-1.8$ yr$^{-1}$, and Gaussian jets $\sim 0.4-1.5$ yr$^{-1}$.}
Note, however that \cite{2013ApJ...764..179B} demonstrated that $\sim 60\%$ of {\it Swift} short GRBs are non-collapsar in origin,
this would reduce the estimated merger rates presented here.

{Here we have considered NS-NS mergers, if short GRBs are from BH-NS mergers only, then the rate will be a factor $\sim10$ larger, where the maximum GW detection distance is approximately twice that for NS-NS mergers.}
As the merger ejecta from a BH-NS is not isotropically distributed, a larger fraction of the ejecta is on the rotational plane, the jet may not propagate through the merger ejecta;
no significant cocoon phase will result in a wider jet.
Any jet structure will be the result of the acceleration/formation mechanism.
The fraction of bright EM jet counterparts to wide homogeneous jets from BH-NS mergers will be higher than those indicated here for NS-NS mergers;
a homogeneous jet with $\theta_j\sim25^\circ$ will produce GRBs in $\sim27\%$ of GW detected mergers, whilst orphan and GRB afterglows with peak flux $m_r\leq21$ will accompany $\la45\%$ of GW detected mergers within the BH-NS GW detection volume $\sim 600$ Mpc.
If the population is all BH-NS mergers with a 25$^\circ$ homogeneous jet, {the merger rate will be $10\leq R_{BH-NS}\leq 40$ yr$^{-1}$, and the number of bright GW-EM counterparts is $4.5-18$ yr$^{-1}$.}
GW-EM counterparts from the jet will be detectable for a significant fraction of BH/NS-NS GW detected mergers;
bright counterparts will typically peak $\la100$ days after the merger.

{Electromagnetic follow-up of a GW trigger requires broadband monitoring of the GW localization region;
a bright optical transient from the jet afterglow, with these models, is expected within $\sim 14$ days.
Optical telescopes with a limiting magnitude of $\sim21$ (e.g. ZTF, Black GEM, GOTO) in joint observations with X-ray and $\gamma$-ray telescopes (e.g. {\it Swift}, {\it Fermi}, MAXI, Chandra) should perform intensive searches/monitoring within the first few weeks.
At later times, any search or monitoring should be conducted by mid- to large-sized telescopes with higher sensitivity (e.g. Subaru HSC, LSST, LT) and radio/infrared observatories (e.g. VLA, ALMA), althogh high-energy monitoring could also reveal a late transient from an off-axis afterglow.
For GW detected mergers that are significantly closer than 200 Mpc, the search timescales should be extended as any transients from structured or off-axis orphan afterglows will be brighter than the limiting detection thresholds for longer.}

{Given one well sampled GW-EM counterpart, the presence of extended jet structure could be revealed if the system is favourably inclined.
An `on-axis', within the jet core angle, afterglow would not reveal any signature of jet structure.
However, afterglows at higher inclinations, or orphan afterglows, could reveal the presence of jet structure;
an achromatic re-brightening would indicate a two-component, or a power-law structured jet.
A shallow decline or slowly brightening afterglow with a soft peak would indicate a Gaussian type jet structure observed at relatively high inclination (within the jet opening angle).
For an off-axis orphan afterglow, either sharp peak followed by a weak decay until a break or a shallow rise to a late peak can be used to indicate the existence of extended jet structure.
Where the prompt emission has been fully suppressed, no X-ray flash or low-luminosity $\gamma$-ray burst, differentiating between an afterglow from within the jet opening angle and a genuine off-axis orphan in the cases of extended jet structure may not be possible.}

\section*{Acknowledgements}
GPL thanks {the annonymous referee, and Y. Fan for useful comments; and} the participants of the Yukawa Institute for Theoretical Physics NPCSM2016 long-term workshop, and S. Nagataki and D. Warren at the Astrophysics Big Bang Laboratory, RIKEN Institute, Japan for useful discussions.
This research was supported by STFC grants; RAS and IAU travel grants.

\bsp
\label{lastpage}
\end{document}